\documentclass{JHEP3}
\usepackage{epsfig,multicol,bbm,latexsym}

\title{Constraint on dark matter annihilation with dark star formation
using Fermi extragalactic diffuse gamma-ray background data}

\author{Qiang Yuan$^{a,b}$, Bin Yue$^{c,d}$, Bing Zhang$^b$ and
Xuelei Chen$^{c,e}$\\
$^a$Key Laboratory of Particle Astrophysics, Institute of High Energy
Physics, Chinese Academy of Sciences, Beijing 100049, P.R. China\\
$^b$Department of Physics and Astronomy, University of Nevada Las Vegas,
Las Vegas, NV 89154, USA\\
$^c$National Astronomical Observatories, Chinese Academy
of Sciences, Beijing 100012, P.R. China\\
$^d$Graduate University of Chinese Academy of
Sciences, Beijing 100049, P.R. China\\
$^e$Center for High Energy Physics, Peking
University, Beijing 100871, P.R. China\\
yuanq@mail.ihep.ac.cn, yuebin@bao.ac.cn, zhang@physics.unlv.edu,
xuelei@cosmology.bao.ac.cn
}

\abstract{
It has been proposed that during the formation of the first generation
stars there might be a ``dark star'' phase in which the power of the
star comes from dark matter annihilation. The adiabatic contraction
process to form the dark star would result in a highly concentrated
density profile of the host halo at the same time, which may give
enhanced indirect detection signals of dark matter. In this work we
investigate the extragalactic $\gamma$-ray background from dark matter
annihilation with such a dark star formation scenario, and employ the
isotropic $\gamma$-ray data from Fermi-LAT to constrain
the model parameters of dark matter. The results suffer from large
uncertainties of both the formation rate of the first generation stars
and the subsequent evolution effects of the host halos of the dark
stars. We find, in the most optimistic case for $\gamma$-ray production
via dark matter annihilation, the expected extragalactic
$\gamma$-ray flux will be enhanced by $1-2$ orders of magnitude.
In such a case, the annihilation cross section of the supersymmetric
dark matter can be constrained to the thermal production level,
and the leptonic dark matter model which is proposed to explain the
positron/electron excesses can be well excluded. Conversely,
if the positron/electron excesses are of a dark matter annihilation
origin, then the early Universe environment is such that no dark
star can form.}


\begin{document}

\section{Introduction}

It has been proposed that there might be a new phase of the stellar
evolution, dark (matter powered) star (DS), during the formation of
the first generation stars (or Pop III stars) in the early time of the
Universe (e.g., \cite{2008PhRvL.100e1101S,2008ApJ...677L...1I,
2008ApJ...685L.101F,2008JCAP...11..014F,2008MNRAS.390.1655I,
2008PhRvD..78l3510T,2009ApJ...692..574N}). Collapse of the gas would
attract dark matter (DM) to collapse together and form a high density
core. Self-annihilation of DM particles would inject energy into the
gas inside the core, which heats the gas core and prevents
further collapse of the gas to initiate nuclear fusion. DSs
typically have distinct features compared with normal stars
\cite{2009ApJ...705.1031S}, e.g., they are massive ($500-1000$
M$_{\odot}$), large ($1-10$ AU), and have low temperature
($T_{\rm surf}<10000$ K). They might be detected with large telescopes
\cite{2010ApJ...716.1397F,2010ApJ...717..257Z}.

The impact of DS formation is very important in many aspects of
astrophysics and cosmology. The reionization history of the Universe
would be altered due to different contribution to the ionization
photons from DSs \cite{2009PhRvD..79d3510S}. The DS remnants might
provide the seeds for super-massive black holes
\cite{2008PhRvL.100e1101S}. Moreover, the formation of DSs would
result in a cuspier density profile of DM outside the stellar
core\footnote{Strictly speaking, it is the adiabatic contraction
which results in a cuspier DM density profile, and due to the same
reason a DS is formed. In the following when we say ``DS
enhancement'' it actually means the enhancement induced by the
contraction process to form DS.}, which is expected to give an
enhanced annihilation luminosity and can be reflected in the
indirect detection signals such as $\gamma$-rays and neutrinos
\cite{2009PhRvD..79d3510S,2011JCAP...01..018S}.

In this work we aim to discuss the contribution of DM annihilation to
the extragalactic $\gamma$-ray background (EGRB), within the framework
of DS formation models. Schleicher et al. have studied the extragalactic
background radiation for the DS scenario and derived the constraints on
DM particle parameters \cite{2009PhRvD..79d3510S}.
There are several improvements or updates in
the current study: 1) A more realistic prescription on the formation
of Pop III stars (which can give birth to the DSs) is adopted;
2) The DM annihilation models are discussed extensively, especially for
the leptonic DM models which may be responsible for the recent reported
electron/positron excesses \cite{2009Natur.458..607A,2008Natur.456..362C,
2008PhRvL.101z1104A,2009A&A...508..561A,2009PhRvL.102r1101A};
3) The EGRB data is updated with the new result released by the
Fermi team \cite{2010PhRvL.104j1101A}, which gives stronger
constraints than the previous EGRET data \cite{1998ApJ...494..523S}.

Throughout the paper a flat $\Lambda$CDM cosmological model is adopted,
and the cosmological parameters are adopted as the results of the combined
analysis of the WMAP 5 year data and other cosmological measurements:
$\Omega_M=0.274$, $\Omega_{\Lambda}=1-\Omega_M$, $\Omega_b=0.046$,
$h=0.705$, $\sigma_8=0.812$, $n_s=0.96$ \cite{2009ApJS..180..330K}.
Adopting a different set of cosmological parameters 
does not alter the final results significantly. For example we have 
checked that the formation rate of first stars would change by less 
than several percent if we adopt the cosmological parameters derived with 
the WMAP 7 year data together with the baryon acoustic oscillation and 
$H_0$ data \cite{2011ApJS..192...18K}.

This paper is organized as follows. In Sect. 2, we present the formation
rate of first generation stars, which is tightly connected with
the formation of DSs. In Sect. 3, we calculate the annihilation luminosity
and $\gamma$-ray flux of both the scenarios with and without DS formation,
and derive the constraints on DM model parameters. The conclusion and
discussion are given in Sect. 4.

\section{Formation rate of the first generation stars}

We first introduce the mass function of the DM halos. The comoving 
number density distribution of DM halos can be expressed as
\begin{equation}
\frac{{\rm d}n(z)}{{\rm d}M}=\frac{\rho_\chi}{M}
\sqrt{\frac{2A^2a^2}{\pi}}\left[1+(a\nu^2)^{-p}\right]\exp\left(-a\nu^2/2
\right)\frac{{\rm d}\nu}{{\rm d}M},
\end{equation}
where $\nu=\delta_c(z)/\sigma(M)$, $\delta_c(z)=1.68/D(z)$
is the critical over-density in spherical collapse model, $D(z)$ is the
linear growth factor \cite{1992ARA&A..30..499C}. $A$, $a$ and $p$ are
constants. For $(A,\,a,\,p)=(0.5,\,1,\,0)$ it is the Press-Schechter (PS)
formula \cite{1974ApJ...187..425P}, and for $(A,\,a,\,p)=(0.322,\,0.707,
\,0.3)$ it is Sheth-Tormen (ST) formula \cite{1999MNRAS.308..119S}.
In this work we adopt the ST mass function.
$\sigma^2(M)$ is the average variance of density field
\begin{equation}
\sigma^2(M)=\frac{1}{2\pi^2}\int W^2(kR_{\rm M})P_\delta(k)
k^2{\rm d}k,
\end{equation}
with the top-hat window function $W(x)=3(\sin{x}-x\cos{x})/x^3$.
$R_{\rm M}=\left(3M/4\pi\rho_m\right)^{1/3}$ is a radius
within which a mass $M$ is contained with a uniform matter
density field. The matter power spectrum $P_\delta(k)$ is expressed as
\begin{equation}
P_\delta(k)=A_s(k\cdot{\rm Mpc})^{n_s}T^2(k),
\end{equation}
where $A_s$ is normalized using $\sigma_8$, transfer function $T(k)$
is obtained from a fit of CDM model \cite{1986ApJ...304...15B},
\begin{equation}
T(q)=\frac{\ln(1+2.34q)}{2.34q}\left[1+3.89q+(16.1q)^2+(5.46q)^3+
(6.71q)^4\right]^{-0.25}
\end{equation}
with $q=k/h\Gamma$ and $\Gamma=\Omega_m h\exp[-\Omega_b(1+\sqrt{2h}/
\Omega_m)]$.

In the early Universe, the first stars would form in 
halos that are massive enough for gas to cool and condense. The 
number of such halos increases rapidly, and the halo destruction by 
mergers can be neglected. We therefore assume that the Star Formation 
Rate (SFR) of the first generation stars is proportional to the redshift 
derivative of the halo mass function $\frac{{\rm d}^2n}{{\rm d}M{\rm d}z}$ 
within a certain mass range \cite{2008ApJ...684...18C}, with a time 
delay $\tau_{d}$ due to the cooling and collapse of the halo 
\cite{2009ApJ...694..879T}. At a given redshift $z$, the SFR can be 
written as
\begin{equation}
\label{sfr}
{\rm SFR_{Pop III}}(z)=\int_z^{\infty}\int_{M_{\rm min}(z'')}
^{M_{\rm max}(z'')}\left |\frac{{\rm d}^2n}{{\rm d}M{\rm d}z}
(M,z'')\right |\left[1-Q_{\rm H^+}(z'')\right] \delta[z-z'(\tau_d,z'')]
{\rm d}M'{\rm d}z'',
\end{equation}
where $Q_{\rm H^+}$ is the volume-filling factor of ${\rm H^+}$ regions,
$z'(\tau_d,z'')$ is the redshift with time $\tau_d$ delay of $z''$.
Here we further assume that only one first generation star could form in 
each potential
halo\footnote{There has been no discussion on whether DSs can form
if fragmentation is important during first star formation (e.g.,
\cite{2009Sci...325..601T}). Our discussion does not apply to those
scenarios.}.
The $\delta$ function in Eq. (\ref{sfr}) means that the halos which
satisfy the formation condition of the first stars will eventually
contribute to the SFR $\tau_d$ later. For the time delay, we adopt
$\tau_d=\tau_{\rm cool}+\tau_{\rm ff}$, in which the cooling
time scale for $\rm H_2$ is $\tau_{\rm cool}=2.38\times10^{13}\left
(\frac{M}{10^6M_\odot}\right)^{-2.627}\left(\frac{1+z}{31}\right)^{-6.94}$ s
and the free fall time scale is $\tau_{\rm ff}=2.77\times10^{14}\left
(\frac{1+z}{31}\right)^{-3/2}$ s \cite{2009ApJ...694..879T}.
This delay time scale $\tau_d$ is mass-dependent, which makes the
calculation of Eq. (\ref{sfr}) more complicated.

As for the lower mass limit, it is usually believed that the first
generation stars could only form in halos with virial temperature above
$10^3$ K if the coolant is $\rm H_2$. However, in the presence of
Lyman-Werner (LW) photons which are emitted by stars which form
previously, this lower limit increases since only $\rm H_2$ in massive
enough halos can survive from the photo-dissociation. Following
Ref. \cite{2009ApJ...694..879T}, the lower limit of the halo mass
$M_{\rm min}$ for the first star formation would be the maximum of
$M_{\rm H_2-cool}=6.44\times10^6M_\odot J_{21}^{0.457}\left(\frac{1+z}{31}
\right)^{-3.557}$ and $M_{\rm vir}(10^3K,z)$, with $J_{21}$ the specific
intensity of LW radiation in unit of $10^{-21}$ erg
s$^{-1}$cm$^{-2}$Hz$^{-1}$sr$^{-1}$.

The upper limit can be adopted as the virial mass corresponding to a
virial temperature of $10^4~K$. It is generally believed that within
halos with $T_{\rm vir} > 10^4~K$, the cooling by atomic Hydrogen
excitation becomes more efficient, so that the gas would fragment into
multiple parts and the massive and isolated first generation stars
cannot form.

The intensity of LW radiation from the first generation stars is
\begin{equation}
\label{jlw}
J_{\rm LW}^{\rm PopIII}(z)=4\pi \int_z^{z_s}n_{\rm PopIII}(z')(1+z')^2
\epsilon_{\rm LW}\frac{{\rm d}l}{{\rm d}z'}{\rm d}z',
\end{equation}
in which $\epsilon_{\rm LW}$ is the emissivity of LW photons,
$\frac{{\rm d}l}{{\rm d}z'}$ is the comoving distance per redshift, $z_s$
is the upper limit above which the LW photons would be redshifted out of
this band, i.e. $z_s=13.6/11.2(1+z)-1$. $n_{\rm PopIII}(z)$ is the
comoving number density of active first generation stars, which can be
obtained through integrating SFR with respect to redshift within the
life time of such stars
\begin{equation}
\label{np}
n_{\rm PopIII}(z)=\int_{z+\Delta z({\tau_{\rm PopIII}})}^{z}
{\rm SFR}_{\rm PopIII}(z') {\rm d}z'.
\end{equation}
We adopt a typical mass of $200$ M$_\odot$ of the first stars in this
work. According to Ref. \cite{2002A&A...382...28S}, the life time of
such a star is $\simeq2.24$ million years, and $\epsilon_{\rm LW}\simeq
1.15\times10^{25}$ erg s$^{-1}$Hz$^{-1}$.
The choice of this mass is consistent with the numerical 
simulations of first stars. If the mass parameter alters by a factor of 
several, the main influence would be the LW feedback. As 
discussed below, we will investigate the LW feedback efficiency within 
a wide range. This effectively includes the variance of the first star 
mass to some extent. We also note that if the initial mass function (IMF)
of the first stars follows a Salpeter form between $100$ M$_\odot$ and 
$500$ M$_\odot$ \cite{2010MNRAS.407.1003M}, the LW emission rate per 
unit star mass is close to that of a $200$ M$_\odot$ star. 

Note that DS formation may also affect the LW feedback. 
First, DSs will also produce LW photons. However, for typical DS 
parameters, $L\sim 10^6-10^7$ M$_{\odot}$, $T\sim 5000-10000$
K \cite{2009ApJ...705.1031S}, we find that the contribution
to LW emissivity ($11.2-13.6$ eV) is negligible. Second, after the
DS stage, the object will enter the main sequence as a traditional star
powered by nuclear fusion, with different properties from the ordinary
PopIII star. There is no clear conclusion about the properties and fate
of this particular main sequence star (e.g., \cite{2009ApJ...705.1031S,
2010ApJ...716.1397F,2010arXiv1006.0025S,2010MNRAS.406.2605R}). In any
case, since we have employed a large range of uncertainty of LW feedback
(see below), the impact of DS formation on the first star formation rate
would not introduce much larger uncertainties, and we have neglected
this effect in the following discussion.

Besides the first generation stars, there are also galaxies that can
contribute to the LW dissociation and ionization process. SFR of
metal-enriched and metal-free galaxies are:
\begin{equation}
{\rm SFR_{gal}^{en}}=p_{\rm en}(z)\times\frac{\Omega_b}{\Omega_m}f_\star
\int_{M_{\rm vir}(10^4{\rm K},z)}^\infty\frac{{\rm d}^2n}
{{\rm d}M{\rm d}z}(M,z){\rm d}M
\label{SFRen}
\end{equation}
and
\begin{equation}
{\rm SFR_{gal}^{fr}}=[1-p_{\rm en}(z)]\times\frac{\Omega_b}{\Omega_m}f_\star
\int_{M_{\rm vir}(10^4{\rm K},z)}^\infty\frac{{\rm d}^2n}
{{\rm d}M{\rm d}z}(M,z){\rm d}M
\label{SFRfr}
\end{equation}
respectively, where $f_\star=2\times10^{-3}$ is the fraction of gas that
converts to stars, and $p_{\rm en}$ is the enrichment probability of gas
by metals. The enrichment probability could be divided into two parts:
the self-enrichment through which a Pop III star could pollute the host
halo, and the probability that a halo is polluted by metals from neighbors.
We follow Ref. \cite{1993MNRAS.262..627L} to calculate the self-enrichment
probability. The pollution probability from neighbors is assumed to be
just the metal fraction of the Universe. According to Ref.
\cite{2002ApJ...567..532H}, almost one half of the mass of a Pop
III star would be released through pair instability supernova (PISN),
and the metal yield from galaxies is about 0.04
\cite{2006MNRAS.373..128G}. As for the LW photons emitted by galaxies,
we take a time-evolved emissivity from Ref. \cite{2003A&A...397..527S},
and add this part to Eq. (\ref{jlw}).

We also trace the reionization evolution history, since it could reduce
the SFR of first generation stars by a factor $1-Q_{\rm H^+}$. The evolution
of $Q_{\rm H^+}$ is described as \cite{2009PhRvD..79d3510S}
\begin{equation}
\frac{{\rm d}Q_{\rm H^+}}{{\rm d}z}=\frac{Q_{\rm H^+}C(z)n_{\rm H^+}
\alpha_A}{H(z)(1+z)}+\frac{{\rm d}n_{\rm ph}/{\rm d}z}{n_{\rm H}},
\label{qh}
\end{equation}
where $C(z)=27.466\exp(-0.114z+0.001328z^2)$ is the clumping factor
\cite{2006MNRAS.372..679M}, $n_{\rm H^+}$ is the number density of
ionized hydrogen, $\alpha_A$ is the recombination coefficient
\cite{1996ApJS..103..467V}, $H(z)$ is the Hubble parameter, and
$n_{\rm H}$ is the mean neutral hydrogen density. In our work, the
production of ionization photons $dn_{\rm ph}/dz$ is expressed as
\begin{equation}
\frac{{\rm d}n_{\rm ph}}{{\rm d}z}(z)=(1+z)^3\left[n_{\rm PopIII}(z)
\epsilon_{\rm ph}+\int_\infty^z {\rm SFR_{gal}}(z')\epsilon_{\rm gal}
(z-z'){\rm d}z'\right]\frac{{\rm d}t}{{\rm d}z}.
\label{dnphdz}
\end{equation}
The first term refers to the contribution from first generation stars.
For a $200$ M$_\odot$ metal-free star the emissivity of ionization
photons $\epsilon_{\rm ph}$ is $3.1\times10^{50}$ s$^{-1}$
\cite{2002A&A...382...28S}. The second term is the contribution from
the galaxies, including metal-enriched and metal-free galaxies. In the
second term, $\epsilon_{\rm gal}$ is a time-evolved emissivity of
ionization photons, which is adopted from Ref. \cite{2003A&A...397..527S}.

There are still large uncertainties about the LW feedback mechanism due to
the lack of understanding of the self-shielding \cite{2003ApJ...592..645Y}.
Furthermore, the molecular hydrogen in the relic HII regions may also
absorb some LW photons, reducing the effect of photo-dissociation.
To consider such a reduction on the LW intensity, we employ a free
parameter $f_{\rm LW}$ to represent the efficiency of LW feedback.

Combining Eqs. (\ref{sfr})\---(\ref{dnphdz}) together, we can get a
coupled equation set of the SFR at a given redshift. This equation set
depends on the evolution history of SFR, so it should be integrated step
by step from an initial time. At a high redshift, e.g. around 60,
$J_{\rm LW}$ is negligible, SFR could be obtained directly
by setting $M_{\rm min}=M_{\rm vir}(10^3{\rm K},z)$. With this initial
value, the SFR at first step after the initial redshift could be calculated
by the set of equations discussed above. Performing the same calculation
step by step, we can finally obtain the entire evolution history of the SFR.

The SFRs of Pop III stars for different LW efficiencies
$f_{\rm LW}=0.0,10^{-4},10^{-2},$ and $1.0$ are shown in Fig.
\ref{fig:SFR}. It is shown that for the very effective LW feedback
the star formation ends very early, at $z\sim 25$. For the most
optimistic case without LW feedback the star formation can last
till $z\sim 6$. These two extreme cases may represent the lower and
upper limits of the SFR of the first generation stars.

\FIGURE{
\includegraphics[width=0.6\columnwidth]{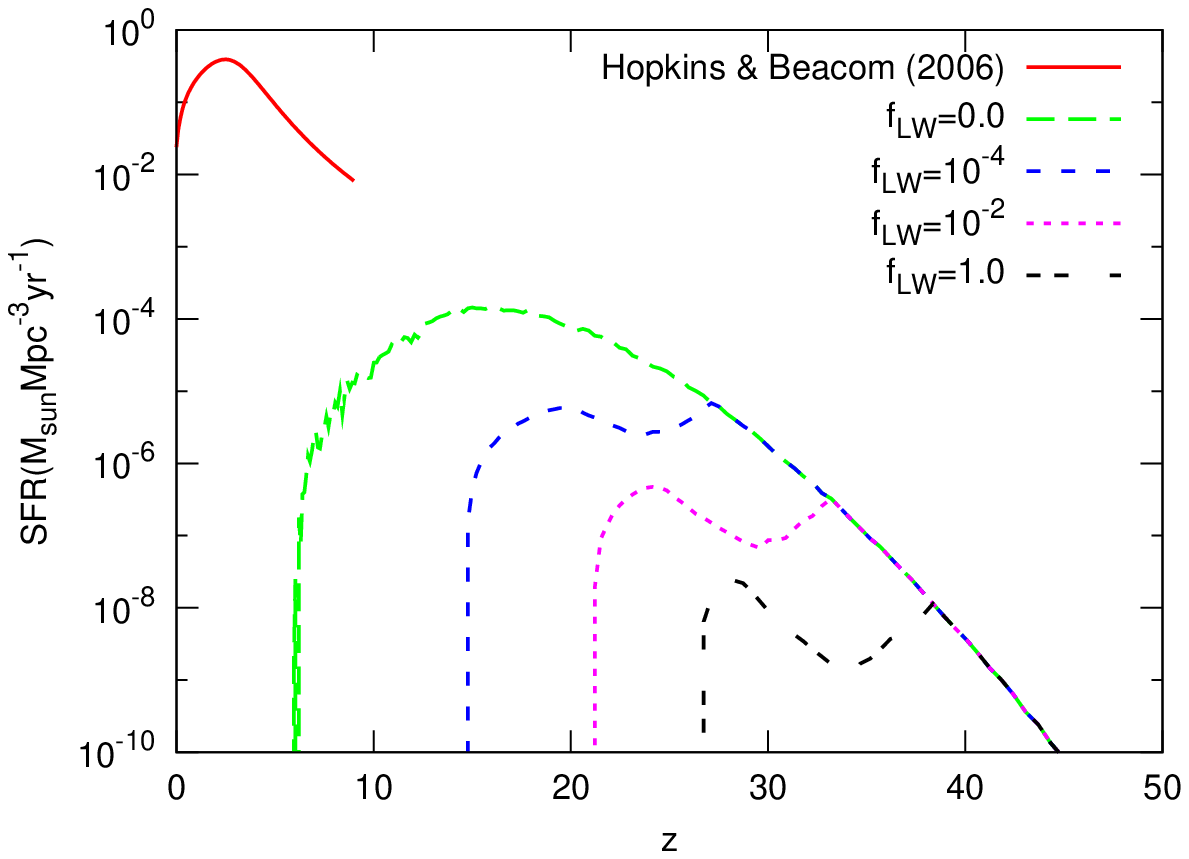}
\caption{The evolution of SFR of first generation stars. In this plot the
SFR is the result of Eq. (\ref{sfr}) multiplied by $M_{\rm fs}
{\rm d}z/{\rm d}t$ to convert to the usual definition. The solid curve
is the fit to the observational SFR for the second and third generation
stars \cite{2006ApJ...651..142H}.}
\label{fig:SFR}
}

\section{EGRB from DM annihilation}

\subsection{Enhancement factor of DM clumpiness}

In this section we describe the enhancement factor of DM annihilation
due to clumpiness, defined as the ratio between DM annihilation rate with
structures and that of smoothly distributed case. We first consider
the standard case without DS formation. The details of the calculation
of the enhancement factor of the extragalactic DM annihilation can be
referred to Refs. \cite{2009PhRvD..80b3517K,2010JCAP...10..023Y}.
Here we quote the basic formulae. The density profile of all the halos
is assumed to be NFW profile \cite{1997ApJ...490..493N}
\begin{equation}
\rho(r)=\frac{\rho_s}{(r/r_s)(1+r/r_s)^2}
\end{equation}
with two scale parameters $r_s$ and $\rho_s$. The scale radius $r_s$
is determined by the concentration parameter $c_v$ of the halo
\begin{equation}
r_s=r_v/c_v,
\end{equation}
where $r_v$ is the virial radius. The scale density $\rho_s$ is then
determined by normalizing the mass of the halo to $M$.
Note that there is no consensus about the density profile
at the innermost part of the halo \cite{1999MNRAS.310.1147M,
2000ApJ...529L..69J,2004MNRAS.349.1039N,2010MNRAS.402...21N,
2006AJ....132.2701G}. For different density profiles of the halos, 
the annihilation signals would
differ from each other very much. However, as shown in 
\cite{2008PhRvL.100e1101S}, the density profiles of the DM halo after
adiabatic contraction are very similar to each other even for very 
distinct initial density profiles. Since in this work we mainly focus 
on the effects of the adiabatic contraction (DS formation), we do not 
discuss in detail the effect of different halo density profiles.

The concentration parameter as a function of halo mass $M$ can be
extracted from N-body simulations. For the low mass halos which are
beyond the resolution of simulations, the concentration is obtained
by extrapolation. In this paper we consider two concentration models. One
is the semi-analytical model developed in Ref. \cite{2001MNRAS.321..559B}
with the update of WMAP5 cosmological parameters, which is labeled as
``B01''. The other model is to extrapolate the power-law fitting results
from the numerical simulations based on the WMAP5 cosmological parameters
\cite{2008MNRAS.391.1940M} directly to the low mass range, which is
labeled as ``power-law''. For both models we employ the redshift
evolution $c_v(z)=c_v(z=0)/(1+z)$ \cite{2001MNRAS.321..559B}.
The results of B01 and power-law models do not differ very much from
each other at high mass scales ($M\gtrsim 10^5$ M$_{\odot}$). However,
when extrapolating to lower mass scales, power-law concentration would
be larger than B01 model, and hence gives larger annihilation signals.
The halos which can give birth to DSs are massive enough that the DS
enhancement between these two concentration models will be very similar.
There are some updated concentration models which show different
mass-dependence and redshift evolution (e.g., \cite{2009ApJ...707..354Z,
2010arXiv1002.3660K}), however, they will also suffer from the
uncertainties when extrapolating to low mass halos.

The total annihilation luminosity for a population of DM halos with
comoving number density distribution ${\rm d}n(z)/{\rm d}M$ is given as
\begin{equation}
L_{\rm tot}(z)=\int{\rm d}M\frac{{\rm d}n(z)}{{\rm d}M}(1+z)^3L(M,z),
\label{Ltot}
\end{equation}
where $(1+z)^3$ is to convert the comoving halo mass function to the
physical one, and $L(M,z) = 4\pi\int_0^{r_v}\rho^2(r)r^2{\rm d}r$ is
the annihilation luminosity of a single halo. For the mass function
${\rm d}n(z)/{\rm d}M$ please refer to the Appendix. The enhancement
factor is then
\begin{equation}
\Delta^2(z)\equiv\frac{L_{\rm tot}}{\rho_{\chi}^2(1+z)^6}~,
\end{equation}
where $\rho_{\chi}$ is the average DM density of the Universe today.

\subsection{DM annihilation luminosity with DS formation}

If there is DS formation following the first generation stars,
the annihilation luminosity will be enhanced. The annihilation
luminosity of the halos which ever contained DSs can be calculated as
\begin{eqnarray}
L_{\rm tot}^{\rm ds}(z)&=&(1+z)^3\int_{z+\Delta z(\tau_{\rm ds})}^z
{\rm d}z' \nonumber \\
&\times & \int_{z'}^{\infty}\int_{M_{\rm min}(z''')}^{M_{\rm max}(z''')}
f_{\rm ds}L(M,z''')[1-Q_{\rm H^+}(z''')]\frac{{\rm d}^2n}{{\rm d}M{\rm d}z'}
\left(M,z'''\right)\nonumber \\
&\times & \delta[z'-z''(\tau_d,z''')]{\rm d}M{\rm d}z''',
\label{Lds}
\end{eqnarray}
where $f_{\rm ds}$ is the enhancement factor of the annihilation
luminosity of a single halo after DS formation compared with the original 
density profile, and $\tau_{\rm ds}$ is the age of the halo which hosts 
a DS. Generally speaking the density profile 
would steepen as $r^{-1.9}$, regardless of the initial density 
profile of the DM halo. We note, however, that the detailed density profile 
would depend on the parameters of the initial DM halo and the collapsing 
gas core. According to the density profile of halos after DS formation 
\cite{2009ApJ...693.1563F}, we estimate $f_{\rm ds}\sim 10^3$ compared
with initial NFW profile. Such a value is consistent with that estimated
in \cite{2009PhRvD..79d3510S}.

\FIGURE{
\includegraphics[width=0.6\columnwidth]{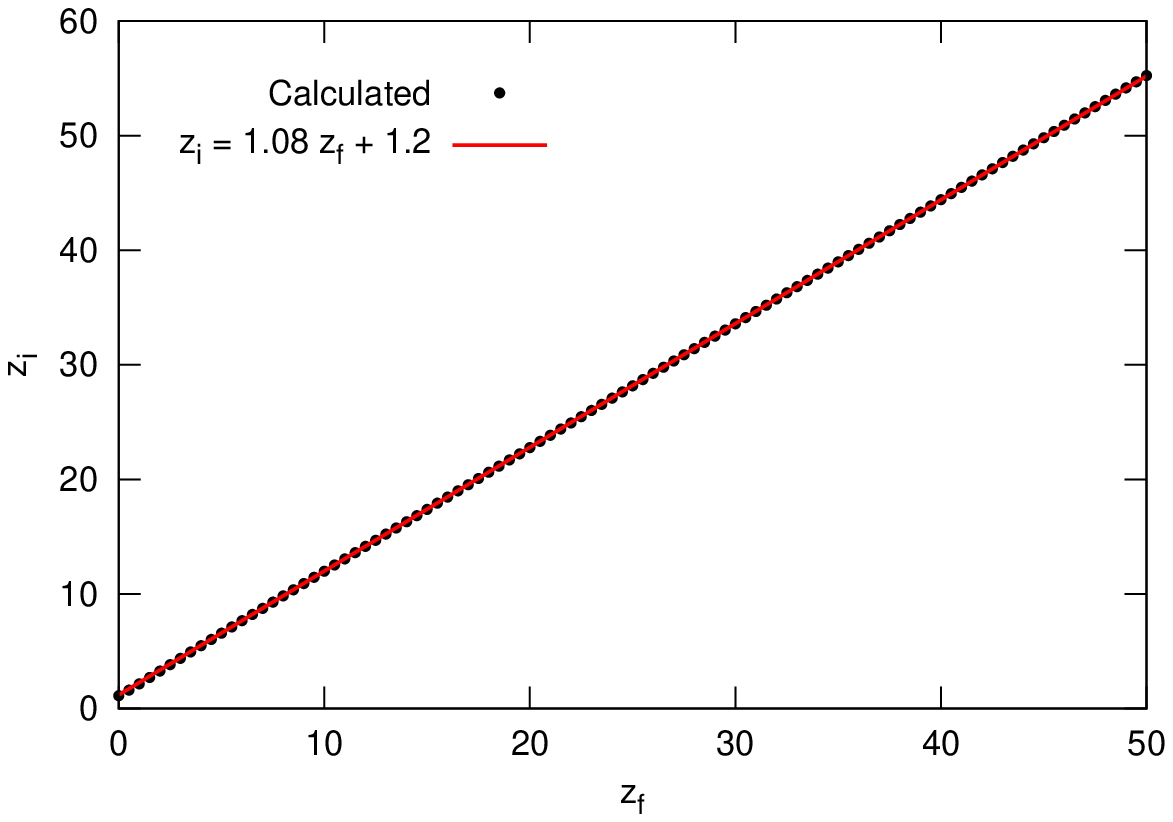}
\caption{Initial redshift $z_i$ of a halo which has a probability
$1-1/e$ when growing from initial mass $M=10^6$ M$_{\odot}$ to $2M$
at $z_f$.
}
\label{fig:zi_zf}
}

If there is only an annihilation effect to consume the DM, $\tau_{\rm ds}$
is estimated to be much longer than the age of the Universe, and the
lower limit of the redshift integral (${\rm d}z'$) can be taken as
$\infty$. On the other hand, in order to qualitatively take into account
the evolution effect of the halos after the DS formation, we also take
another conservative approach by considering that the halo with DS
formation can only exist for a finite time due to possible mergers
with other halos\footnote{According to the numerical simulations of Ref.
\cite{2004MNRAS.349.1117B}, the result of a merger between two halos
with different density profiles could attain a density profile with an
intermediate cusp slope. However, in Ref. \cite{2006ApJ...641..647K}
it was found that the final density profile would be essentially close
to the cuspier one. Here we assume the initial density cusp $r^{-1.9}$
of the dark-star-enhanced halos gradually shallows due to the mergers
with more abundant NFW halos. This gives a conservative bound of
the evolutionary effect of the halo density profiles.}.
Following Ref. \cite{1993MNRAS.262..627L}, for a halo with mass
$M\simeq 10^6$ M$_{\odot}$ at redshift $z_i$, we define a surving time
$\tau_{\rm ds}$ (corresponding to a final redshift $z_f$) after which
the halo reaches a mass $2M$ due to merger or accretion. There will be
a probability distribution of the final redshift $p(z_f,2M;z_i,M)$.
The characteristic final redshift is defined as $P(z>z_f,2M;z_i,M)=1-1/e$,
where $P$ is the accumulative probability that the halo reaches $2M$
before $z_f$. Using the probability distribution Eq. (2.22) in
Ref. \cite{1993MNRAS.262..627L} we find that an empirical fit
$z_i\approx 1.08 z_f+1.2$ nicely applies for
the cosmological model adopted in this work, as shown in Fig.
\ref{fig:zi_zf}. As a result, the lower limit of redshift integration in
Eq. (\ref{Lds}) is simply taken as $1.08 z+1.2$.
These two cases (infinite and minimum ages) bracket the possible
scenarios of DS evolution history.

Replacing $L_{\rm tot}$ in Eq. (\ref{Ltot}) with
$L_{\rm tot}^{\rm ds}$ we can derive the enhancement factor
$\Delta_{\rm ds}^2$ for the halos with DS formation. The
enhancement factors are shown in Fig. \ref{fig:bz}. It is shown that
for the most optimistic case, i.e., without LW feedback and no evolution
of the DS host halos, the enhancement factor due to DM clumpiness
is found to be larger by $1-2$ orders of magnitude compared with the
scenario without DS formation. However, for most other cases the additional
enhancement effect due to DS formation is negligible.

\FIGURE{
\includegraphics[width=0.45\columnwidth]{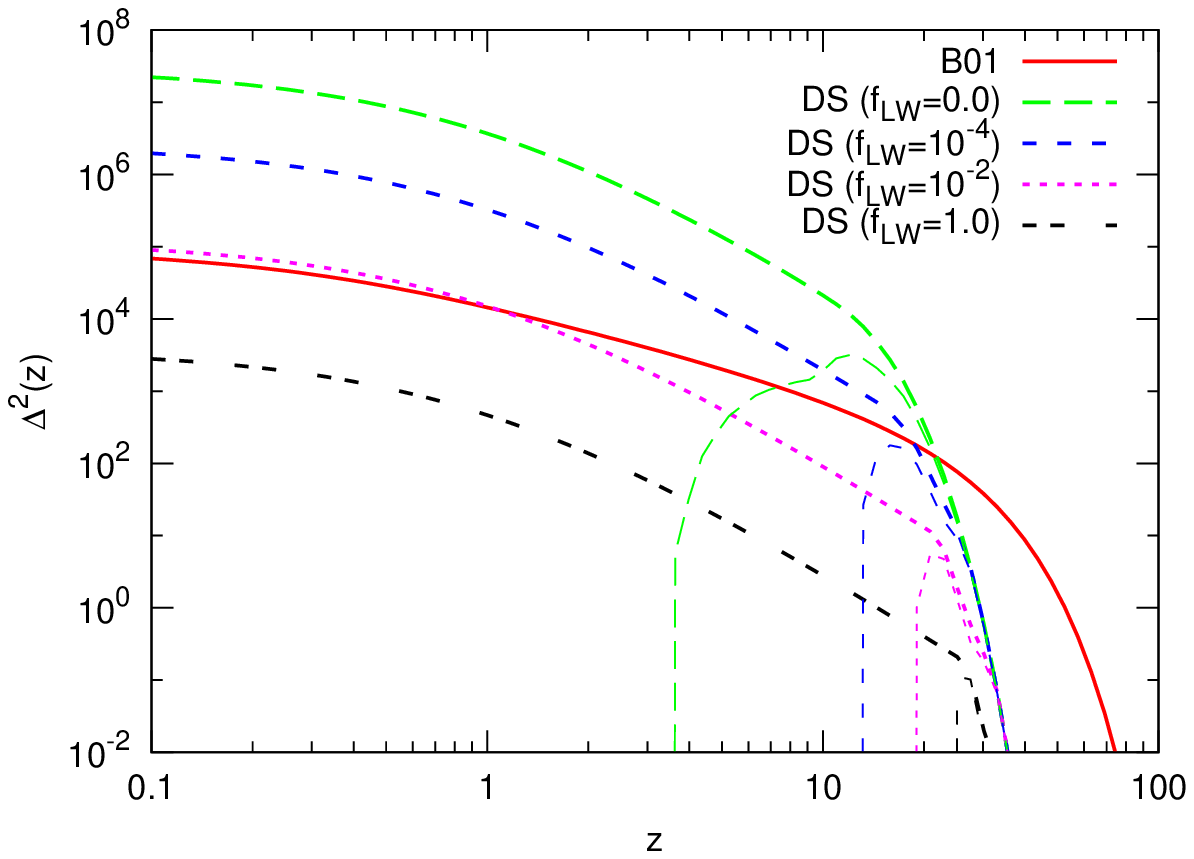}
\includegraphics[width=0.45\columnwidth]{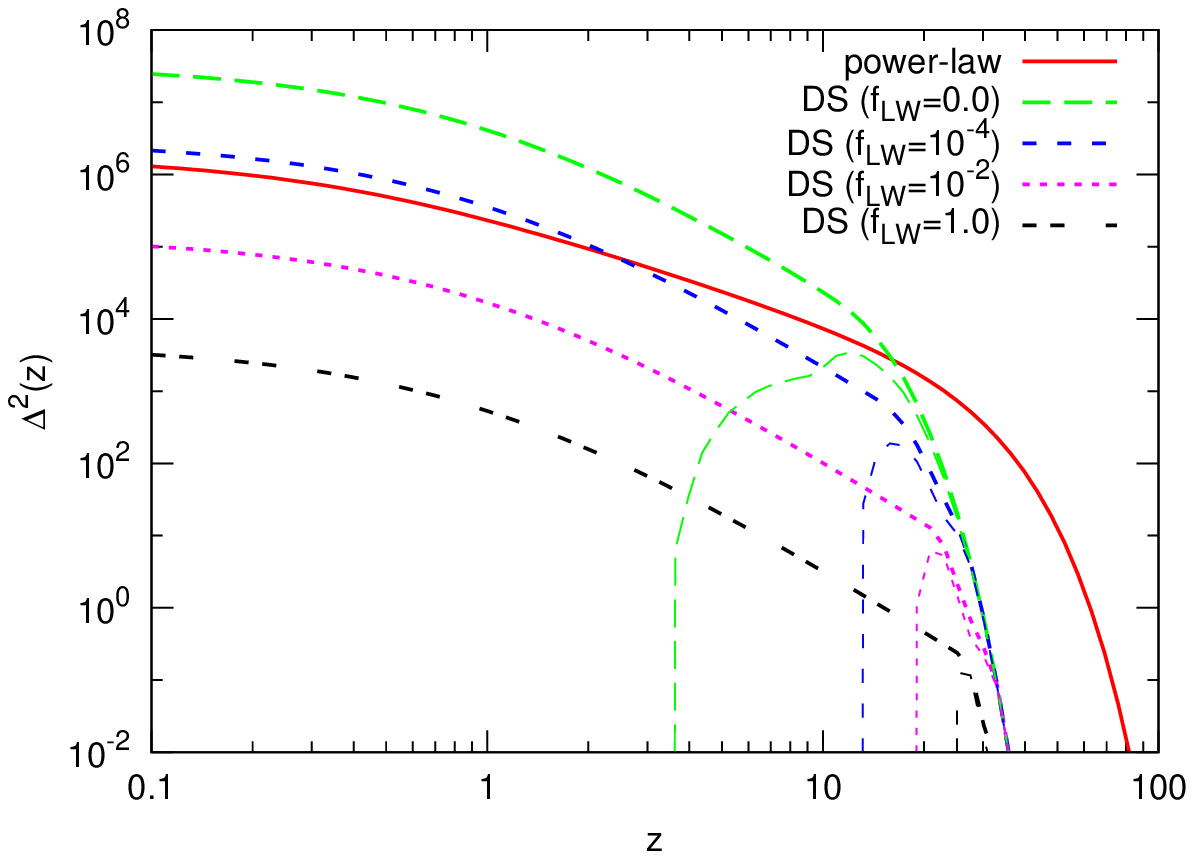}
\caption{Enhancement factor of DM annihilation from clumpiness. Two
concentration models, B01 (left) and power-law (right), are adopted.
The red solid line is for the case without DS formation. Other lines
are the enhancement factors for the halos with DS formed, for infinite
(thick) and minimum (thin) age of the halos. From top to bottom the
curves represent different LW efficiencies $f_{\rm LW}=0.0,10^{-4},
10^{-2},1.0$ respectively.
}
\label{fig:bz}
}

\subsection{Gamma-ray flux and constraints on DM model parameters}

The $\gamma$-ray flux produced by DM annihilation observed today
can be derived through integrating the emissivity over the
evolution history of the Universe
\cite{2002PhRvD..66l3502U,2009A&A...505..999H},
\begin{equation}
\phi(E)=\frac{c}{4\pi}
\frac{\Omega_\chi^2\rho_c^2\langle\sigma v\rangle}{2m_\chi^2}
\int_0^{\infty}{\rm d}z'\frac{(1+z')^3[1+\Delta^2(z')]}{H(z')}
\frac{{\rm d}N}{{\rm d}E'}\exp\left[-\tau(z',E')\right], \label{phi}
\end{equation}
where $m_\chi$ is the mass of DM particle, $\langle\sigma v\rangle$
is the velocity weighted annihilation cross section, $\Omega_{\chi}
\approx 0.23$ is the DM density parameter, $\rho_c=3H_0^2/8\pi G$
is the critical density of the Universe at present, $H(z)$ is the
Hubble parameter, $E'\equiv E(1+z')$, ${\rm d}N/{\rm d}E'$
is the $\gamma$-ray generation multiplicity at redshift $z'$ for
one annihilation of a pair of DM particles, and $\tau(z',E')$ is
the optical depth of the $\gamma$-ray photons when propagating in
the intergalactic space.

The $\gamma$-ray photons of both the primary component which are
generated directly from the DM annihilation products and the secondary
inverse-Compton (IC) component due to scattering of
DM-induced electrons/positrons off the cosmic microwave background
(CMB) photons are included. The photon spectrum of the primary
component for specified annihilation mode is calculated using the
simulation code PYTHIA \cite{2006JHEP...05..026S}. For the secondary
IC component, we assume the electrons and positrons will cool
instantaneously after the production \cite{2009JCAP...07..020P}.
Then the equilibrium $e^{\pm}$ spectrum ${\rm d}n_e/{\rm d}E_e$ is
simply the solution of the energy loss equation
$$
-\frac{\partial}{\partial E_e}\left[\frac{{\rm d}E_e}{{\rm d}t}
\frac{{\rm d}n_e}{{\rm d}E_e}\right]+\frac{{\rm d}N_e}{{\rm d}E_e}=0,
$$
where the energy loss rate ${\rm d}E_e/{\rm d}t=2.67\times 10^{-17}
(1+z)^4(E_e/{\rm GeV})^2$ GeV s$^{-1}$, ${\rm d}N_e/{\rm d}E_e$ is
the production $e^{\pm}$ multiplicity per annihilation. Then the IC
photon spectrum can be calculated through convolving the $e^{\pm}$
spectrum with the CMB photon spectrum and the Klein-Nishina
differential cross section \cite{1970RvMP...42..237B}.

The processes absorbing $\gamma$-ray photons include photo-ionization,
photon-nuclei pair production, Compton scattering, photon-photon
scattering and photon-photon pair production \cite{1989ApJ...344..551Z,
2004PhRvD..70d3502C}. For redshift lower than $6$, we also consider the
pair production when scattering off the cosmic infrared background,
adopting the ``baseline'' model in Ref. \cite{2006ApJ...648..774S}.

\FIGURE{
\includegraphics[width=0.45\columnwidth]{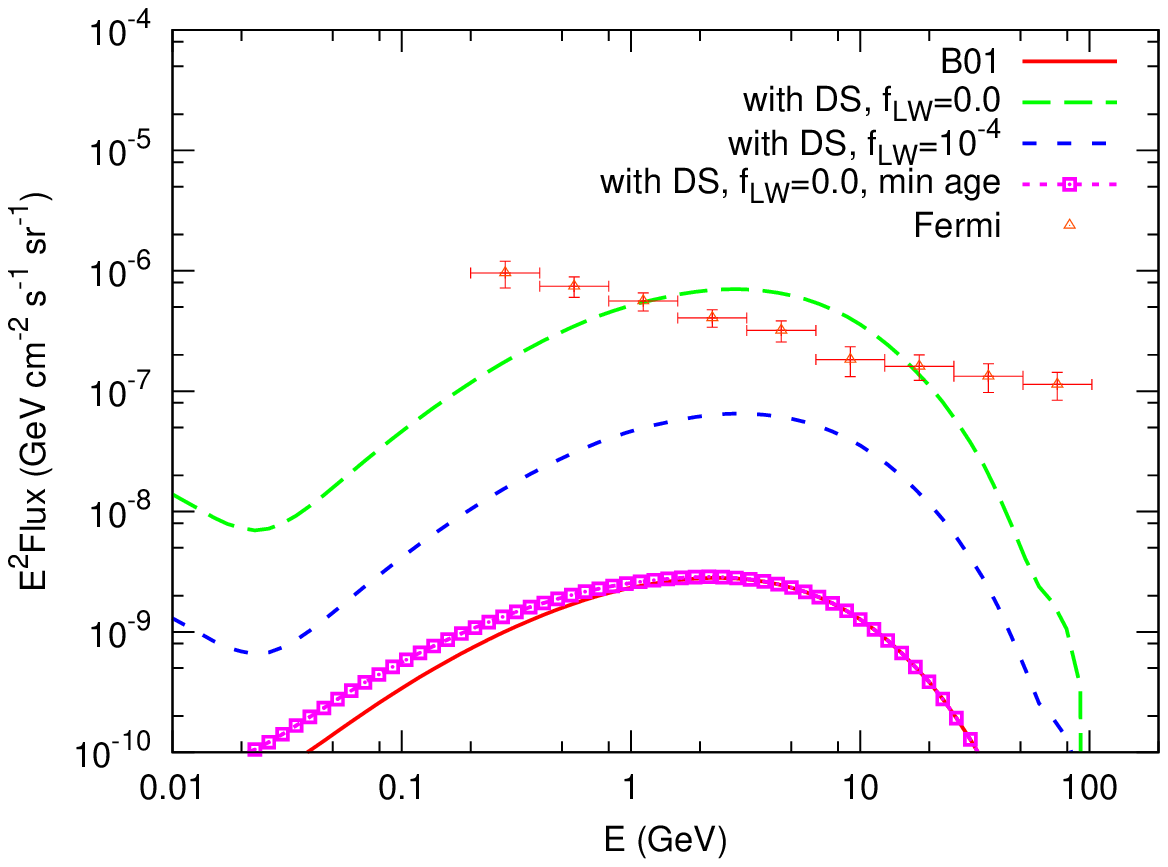}
\includegraphics[width=0.45\columnwidth]{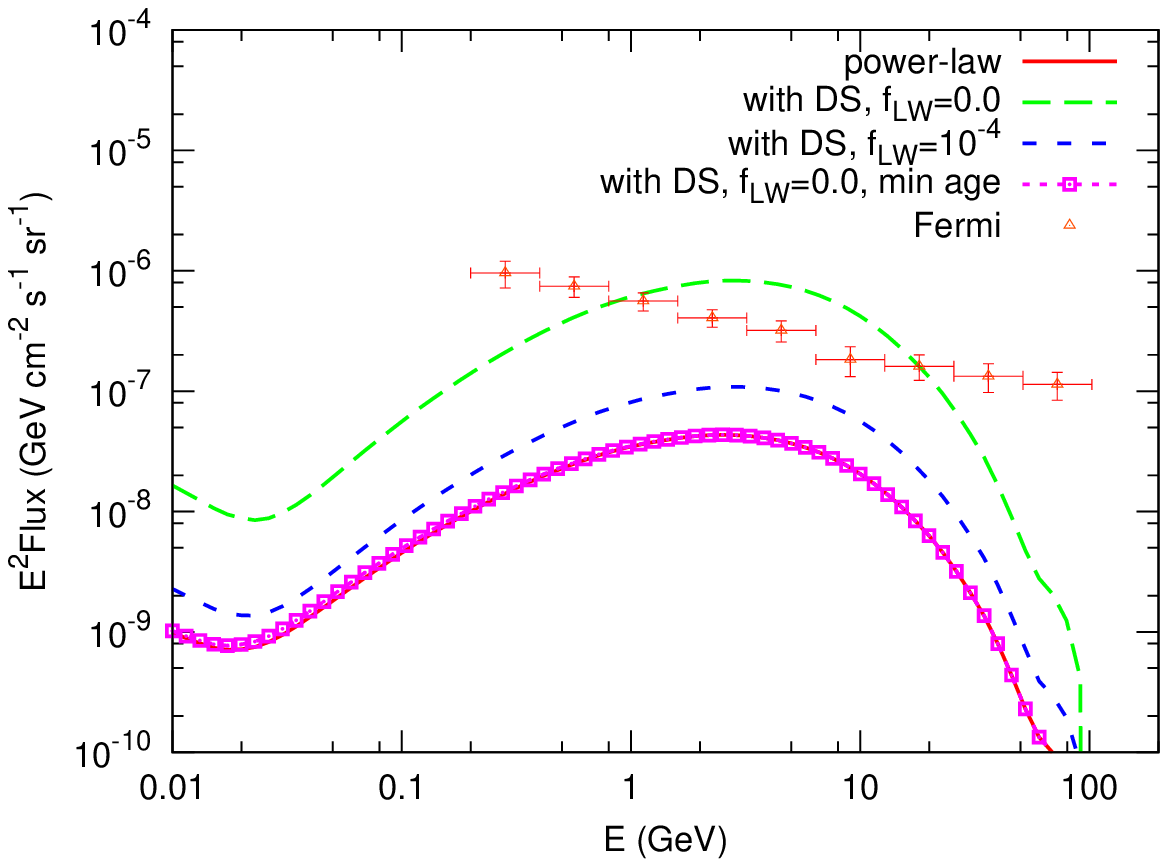}
\includegraphics[width=0.45\columnwidth]{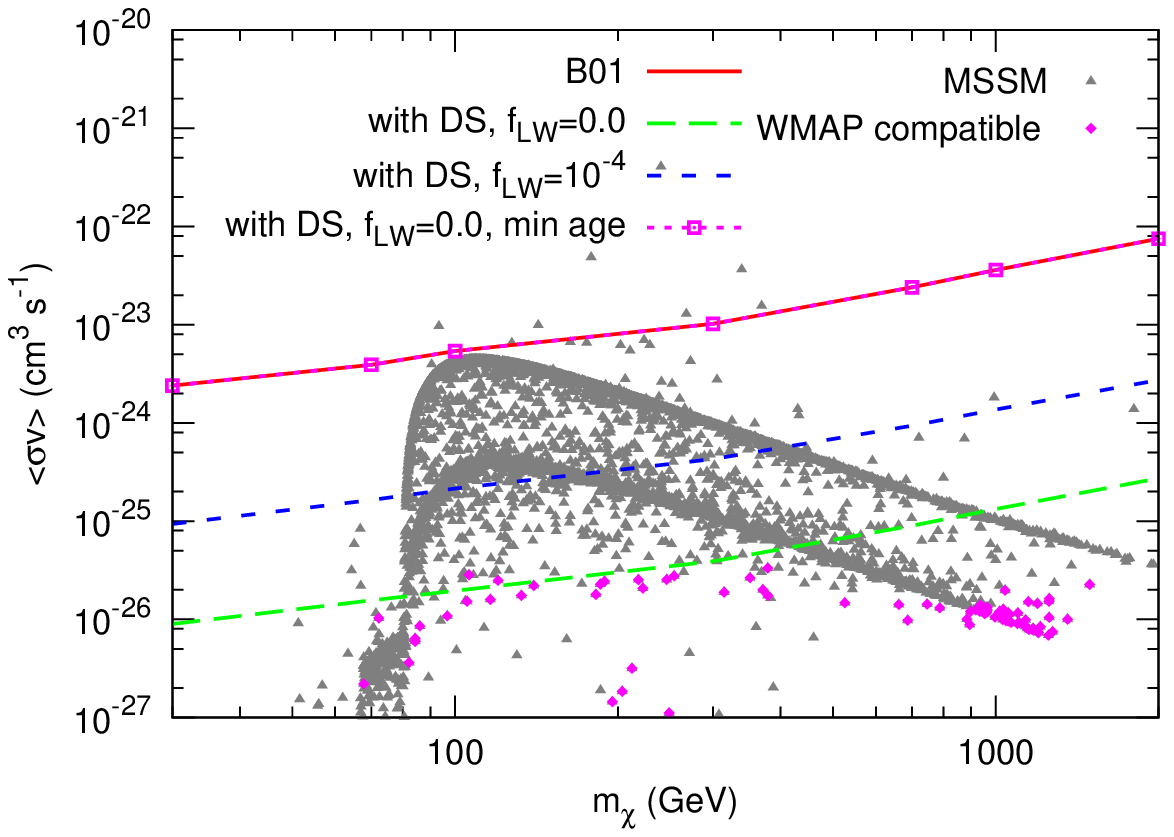}
\includegraphics[width=0.45\columnwidth]{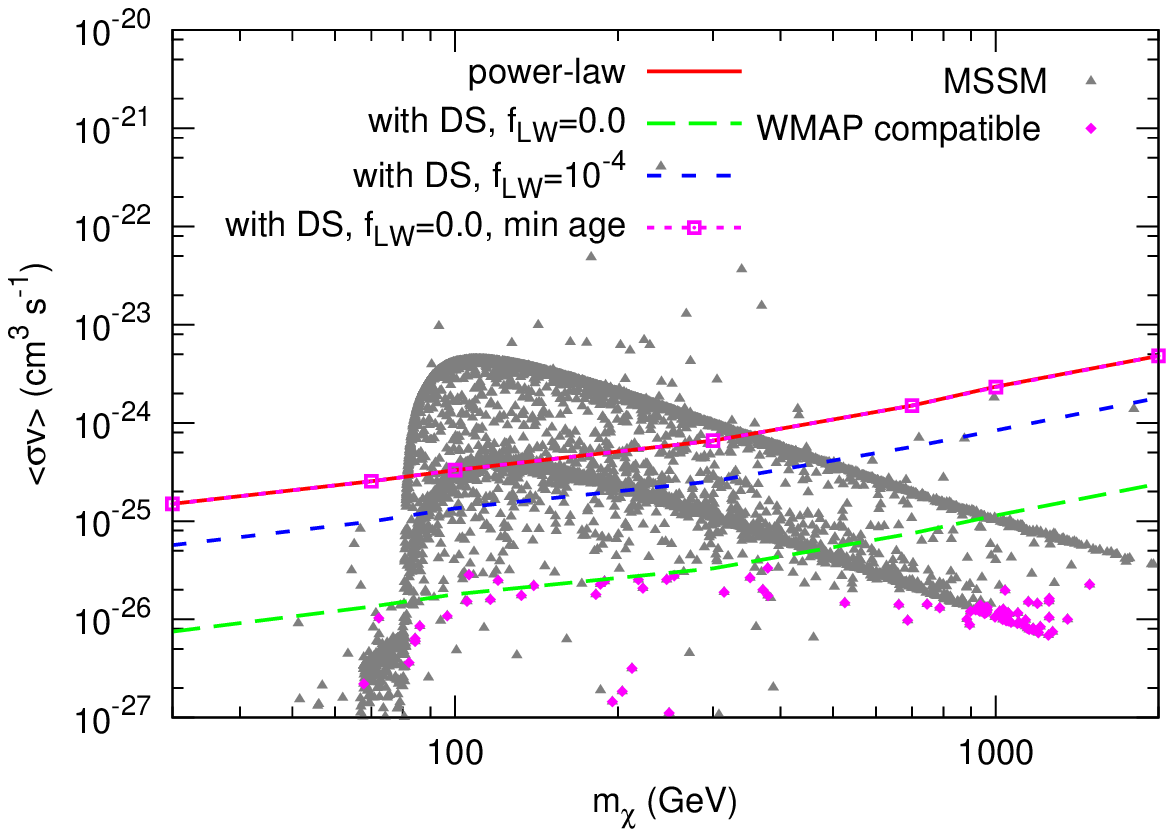}
\caption{Upper panels: comparison of the expected EGRB fluxes from DM
annihilation with the observational data by Fermi
\cite{2010PhRvL.104j1101A}. The particle parameters of DM are adopted
as the canonical neutralino-like WIMP models with $m_\chi=100$ GeV,
$\langle\sigma v\rangle=3\times 10^{-26}$ cm$^{-3}$ s$^{-1}$, and $b\bar{b}$
annihilation final state. Lower panels: constraints on the $m_{\chi}-
\langle\sigma v\rangle$ parameter plane of the DM (regions below the
lines are allowed). The dots show the random scan of the MSSM
model parameters with DarkSUSY \cite{2004JCAP...07..008G}, with relic
density of DM compatible with (diamond) or lower than (triangle) WMAP
observational result. The left and right panels are for B01 and power-law
concentrations respectively.
}
\label{fig:eg_bb}
}

We first consider the canonical neutralino-like DM model. The expected
EGRB fluxes for $m_{\chi}=100$ GeV, $\langle\sigma v\rangle=3\times
10^{-26}$ cm$^3$ s$^{-1}$, $b\bar{b}$ final state are shown in the
upper panels of Fig. \ref{fig:eg_bb}. The left panel is for B01
concentration model, and the right one is for power-law concentration.
The solid line denotes the ordinary model without DS formation. Several DS
models are plotted to show the enhancement effect of DS formation:
$f_{\rm LW}=0,\,\tau_{\rm ds}=\infty$ (long dashed); $f_{\rm LW}=10^{-4},\,
\tau_{\rm ds}=\infty$ (short dashed); $f_{\rm LW}=0$ with minimum
$\tau_{\rm ds}$ (dotted). We can see that for the case with the strongest
DS enhancement the expected EGRB will exceed the Fermi data and should be
excluded. On the other hand, for the models without DS formation or
models with moderate DS enhancement, the data can not give very
strong constraint.

The exclusion limits on the $m_{\chi}-\langle\sigma v\rangle$ plane,
above which the parameter space is excluded, are shown in the lower
panels of Fig. \ref{fig:eg_bb}. Here the constraints
are derived by requiring that the EGRB not to exceed the $2\sigma$
errorbars of the Fermi data, like the conservative method given in
\cite{2010JCAP...04..014A}. Actually since the spectral shape of Fermi
data is very different from that expected from DM annihilation, we can
assume a power-law background and expect to give a much stronger
constraint on the DM contribution \cite{2010JCAP...04..014A}.
For comparison we also give the supersymmetric model predicted
parameters through a random scan of the parameter space of the
Minimal Supersymmetric extension of the Standard Model (MSSM), using
DarkSUSY package \cite{2004JCAP...07..008G}. It is shown that for the
model with the largest DS enhancement, a large part of the MSSM parameter
space can be excluded. The exclusion limits even reach the range required
by the relic density of DM if DM is produced thermally in the early
Universe.

\FIGURE{
\includegraphics[width=0.45\columnwidth]{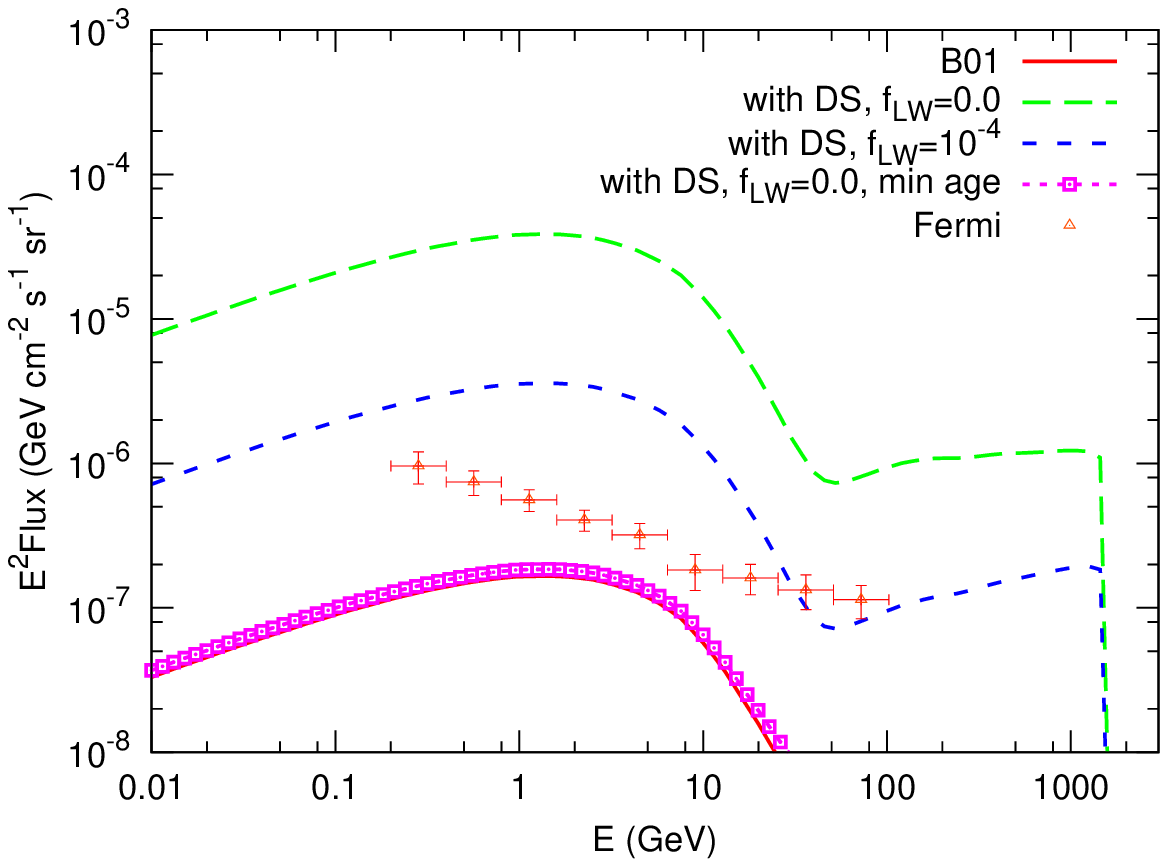}
\includegraphics[width=0.45\columnwidth]{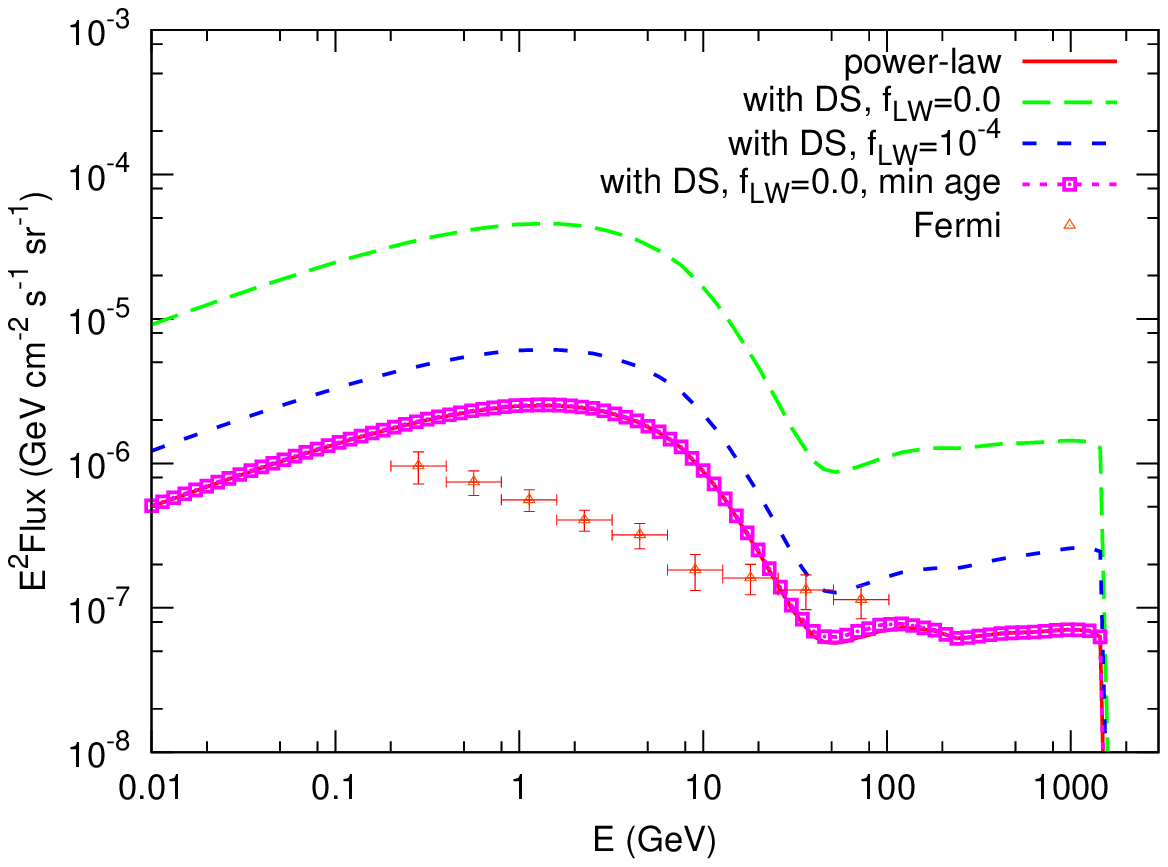}
\includegraphics[width=0.45\columnwidth]{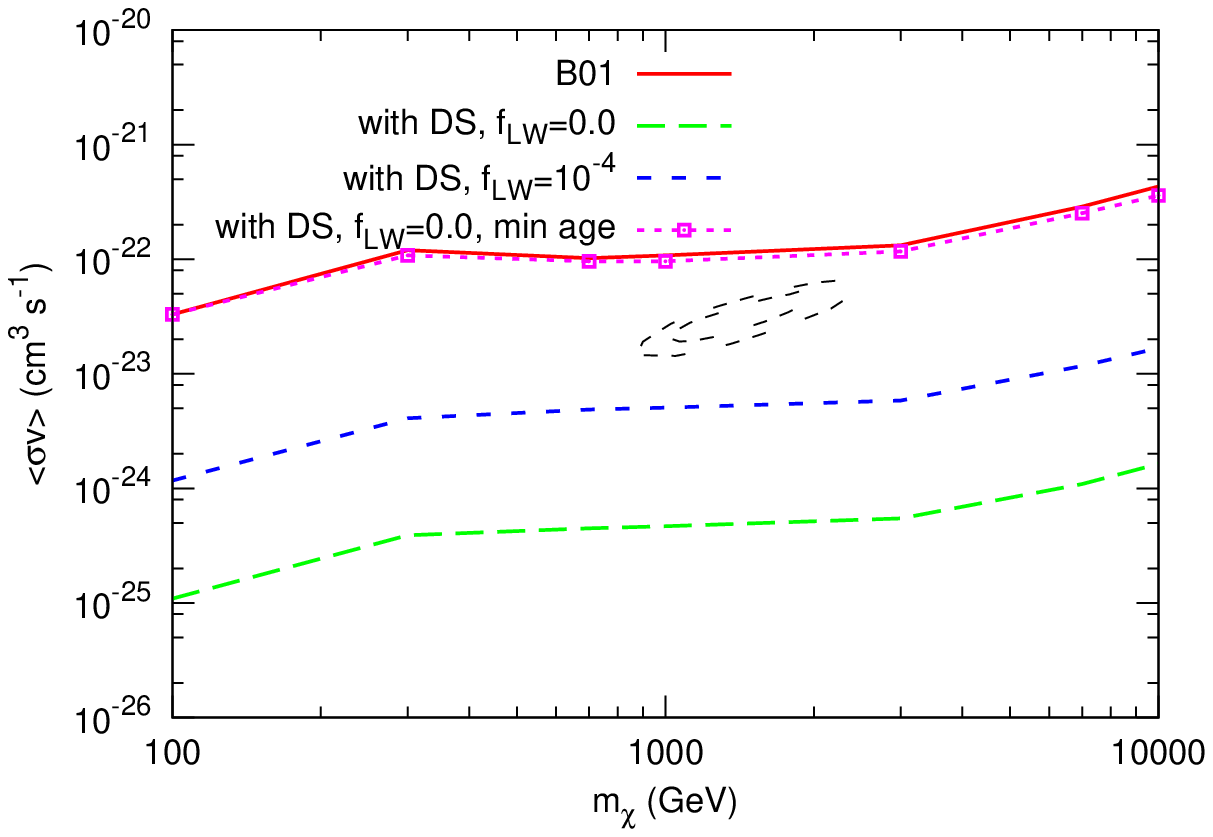}
\includegraphics[width=0.45\columnwidth]{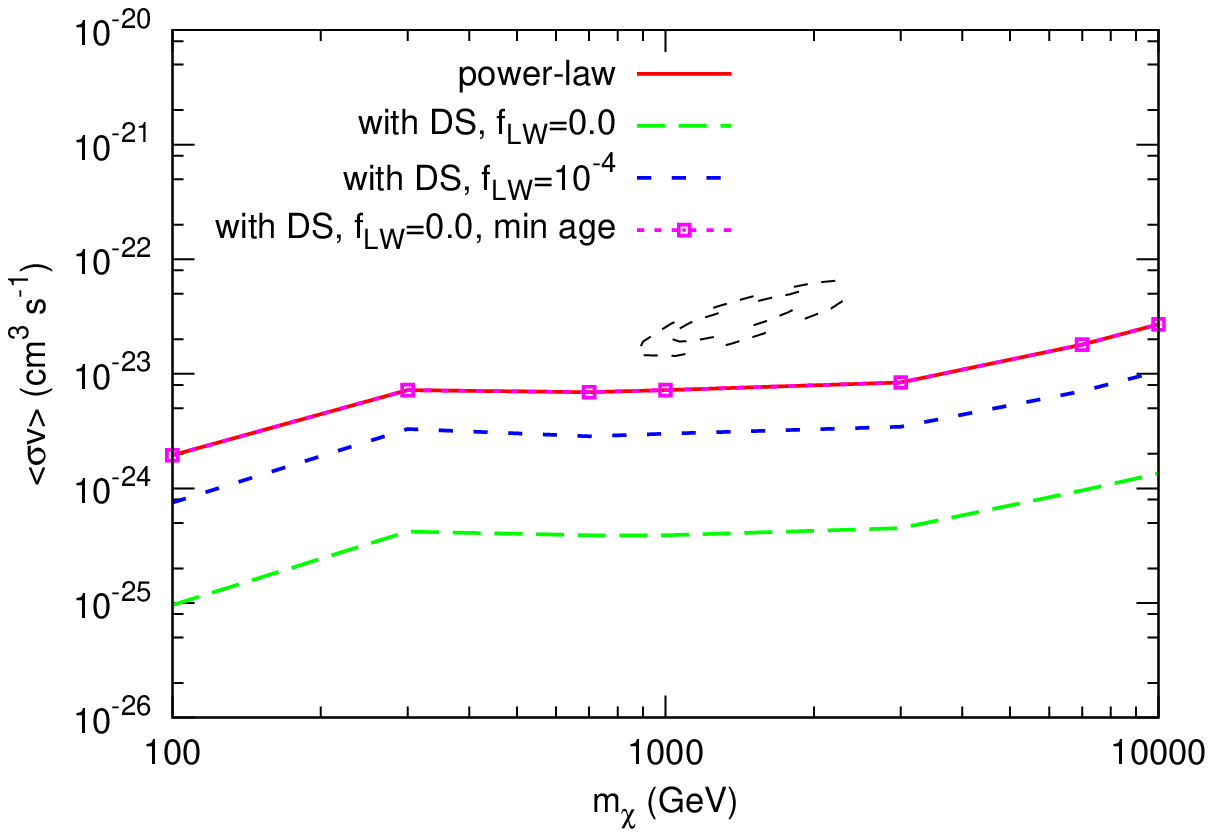}
\caption{Same as Fig. \ref{fig:eg_bb} but for leptonic DM model
with $\mu^+\mu^-$ annihilation final state, which is proposed to be
able to explain the recent observed electron/positron excesses.
The mass of DM is $m_\chi=1.7$ TeV, and the annihilation cross section
is $\langle\sigma v\rangle=3.6\times 10^{-23}$ cm$^{-3}$ s$^{-1}$.
In the lower panels the contours are $3\sigma$ and $5\sigma$ confidence
regions of the fit to PAMELA/Fermi/HESS electron and positron data
\cite{2010NuPhB.831..178M}.
}
\label{fig:eg_mu}
}

\FIGURE{
\includegraphics[width=0.45\columnwidth]{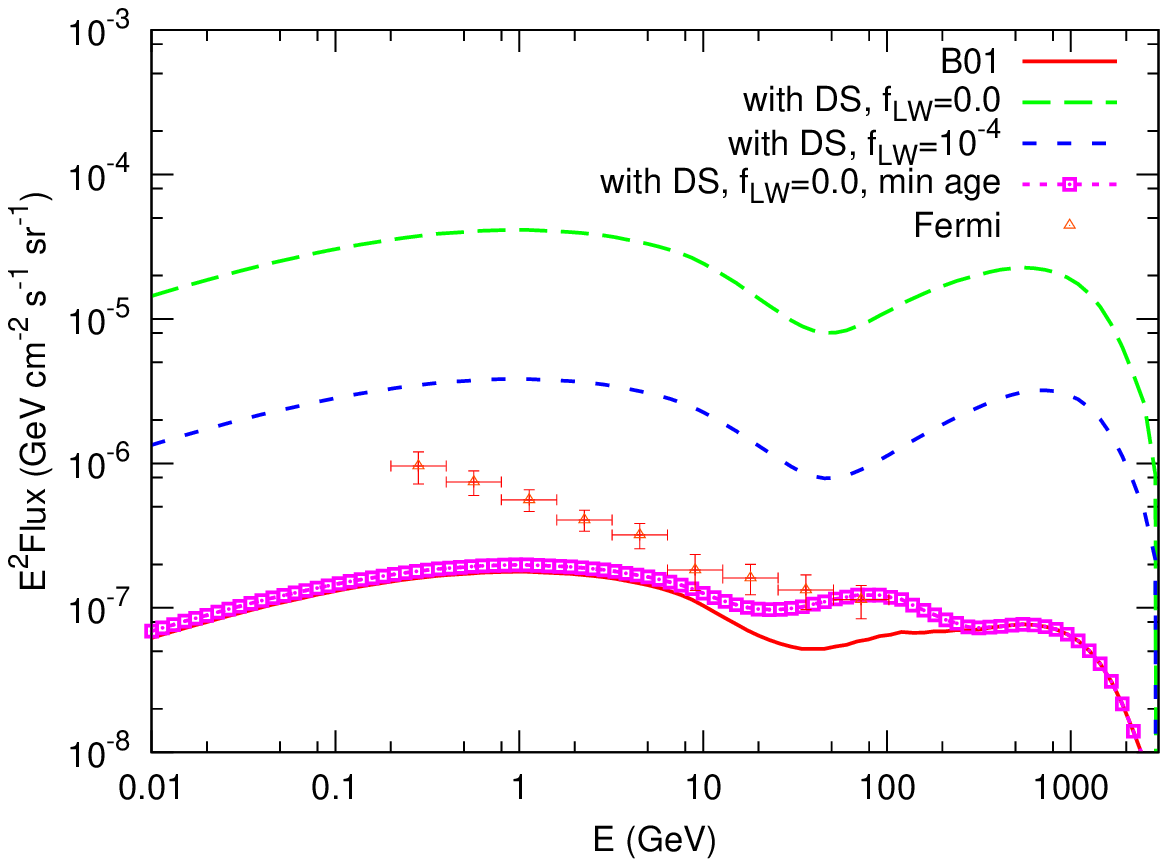}
\includegraphics[width=0.45\columnwidth]{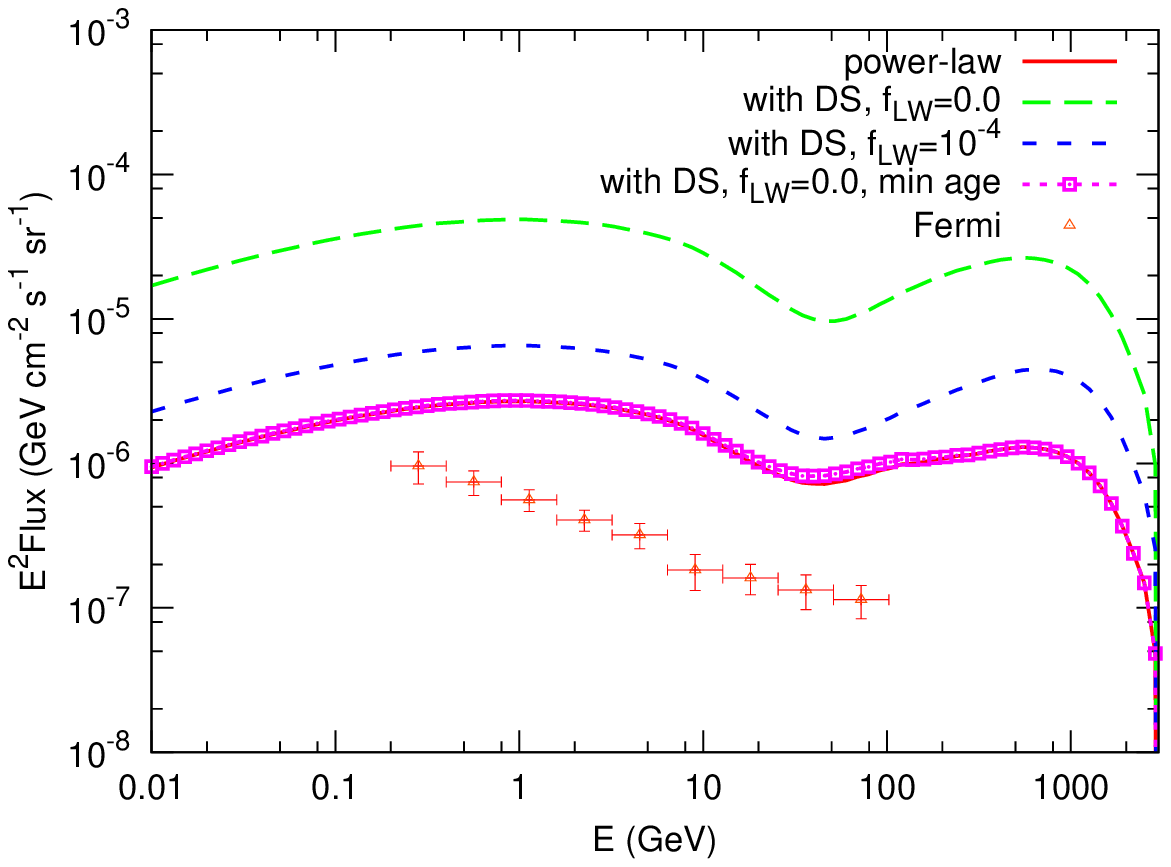}
\includegraphics[width=0.45\columnwidth]{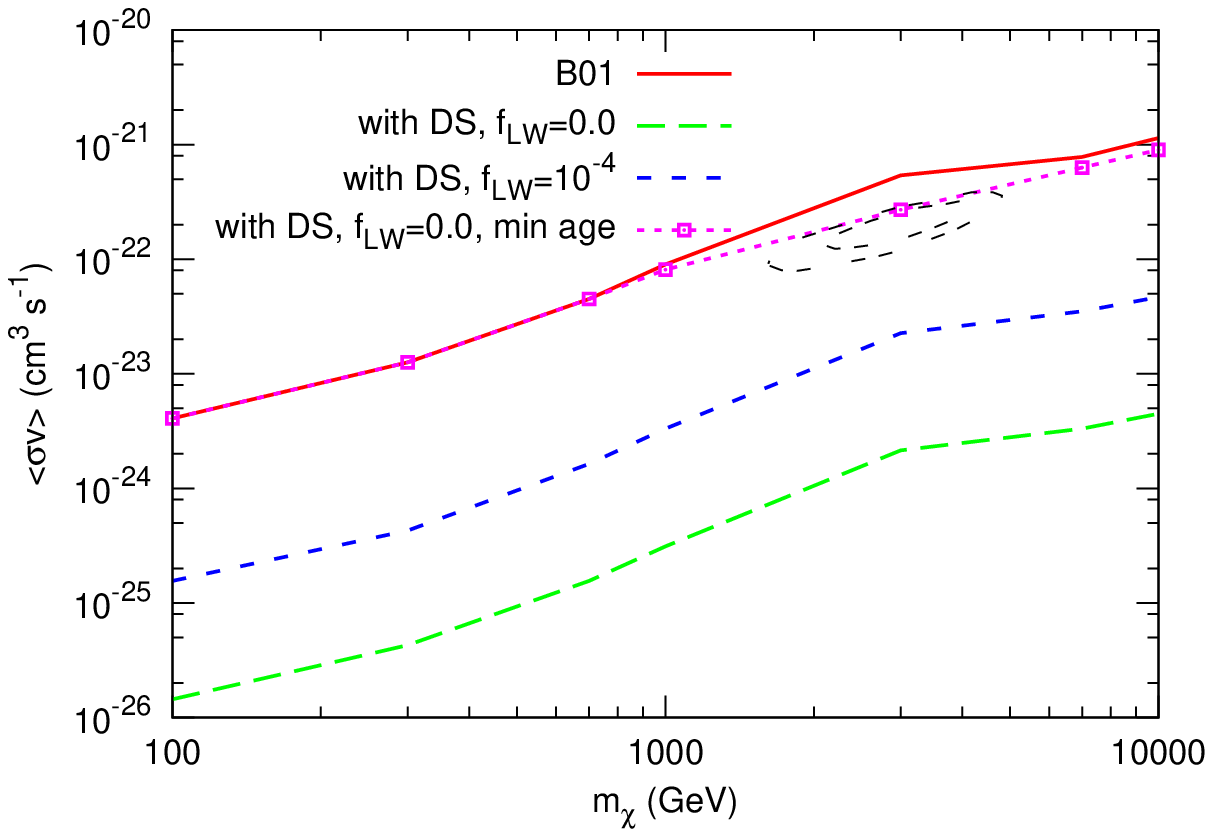}
\includegraphics[width=0.45\columnwidth]{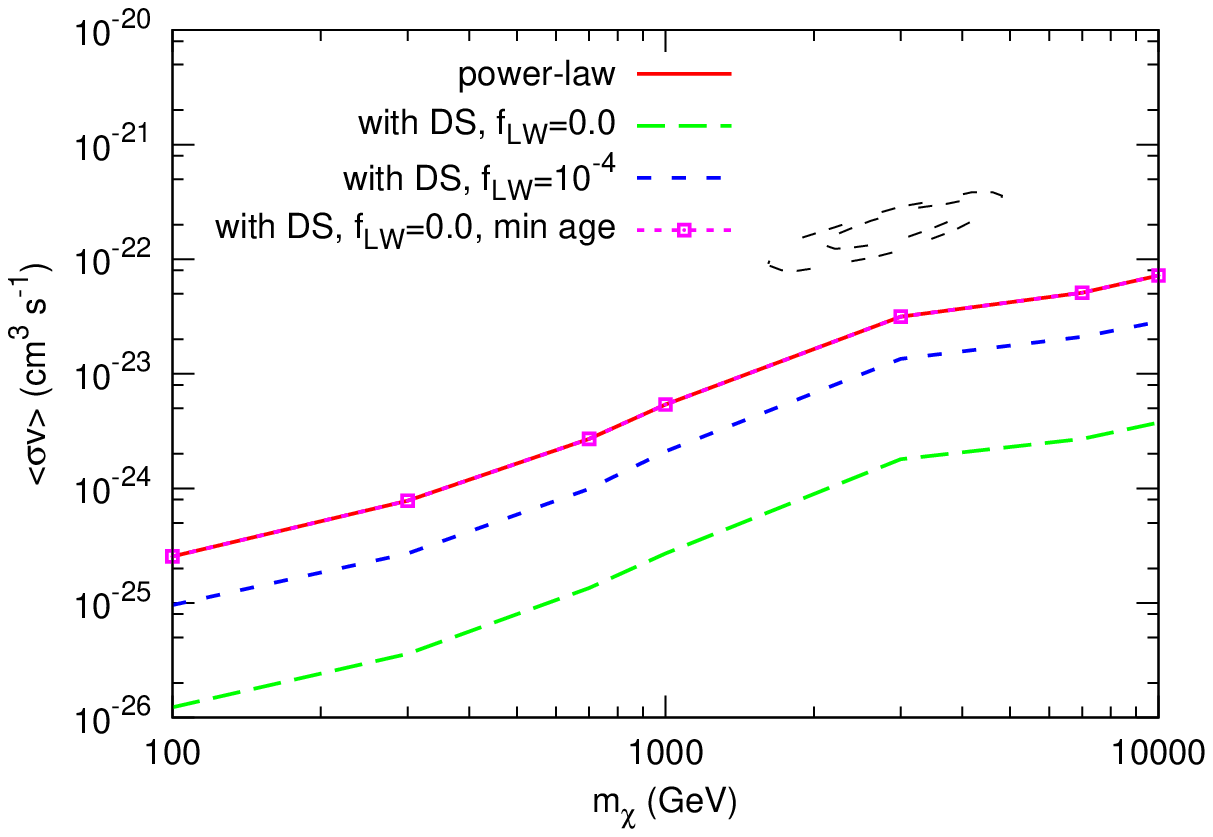}
\caption{Same as Fig. \ref{fig:eg_mu} but for leptonic DM model
with $\tau^+\tau^-$ annihilation final state. The mass of DM is
$m_\chi=3$ TeV, and the annihilation cross section is
$\langle\sigma v\rangle=1.9\times 10^{-22}$ cm$^{-3}$ s$^{-1}$.
}
\label{fig:eg_tau}
}

What of most interest are the recent discoveries of the high energy
electron/positron excesses by PAMELA \cite{2009Natur.458..607A},
ATIC \cite{2008Natur.456..362C}, HESS \cite{2008PhRvL.101z1104A,
2009A&A...508..561A} and Fermi \cite{2009PhRvL.102r1101A}. Together
with the non-excess of antiproron-proton ratio
\cite{2009PhRvL.102e1101A}, the leptonic DM model is favored if DM
is responsible for the excesses (e.g., \cite{2008PhRvD..78j3520B,
2009NuPhB.813....1C,2009PhRvD..79b3512Y,2009PhRvD..80l3518C,
2009PhRvD..79a5014A,2009PhRvL.103c1103B,2010PhLB..691...18H,
2010NuPhB.831..178M,2010NuPhB.831..217P,2010JHEP...12..006I}). 
Alternatively, many astrophysical models have been proposed to interpret 
the signature (see \cite{2010IJMPD..19.2011F} for a review). In this work
we consider the DM annihilating to $\mu^+\mu^-$ or $\tau^+\tau^-$ final
states, which are proposed to fit the PAMELA/Fermi/HESS data of positrons
and electrons.

The results for $\mu^+\mu^-$ final state are presented in Fig.
\ref{fig:eg_mu}. In the upper panels, the illustrating EGRB fluxes for
$m_{\chi}=1.7$ TeV and $\langle\sigma v\rangle=3.6\times 10^{-23}$
cm$^{-3}$ s$^{-1}$ are shown. The lower panels give the exclusion limits
on the $m_{\chi}-\langle\sigma v\rangle$ plane. The dashed contours in
the lower panels are the $3\sigma$ and $5\sigma$ confidence
regions of the fit to PAMELA/Fermi/HESS electron and positron data
\cite{2010NuPhB.831..178M}. We can see that even for the case without
DS enhancement, the $e^{\pm}$ favored parameter regions are excluded for
power-law concentration model by the Fermi EGRB data. For B01 concentration
model the exclusion limit is also very close to the $e^{\pm}$ favored
region\footnote{Compared with the similar semi-analytical model
``BulSub'' of Ref. \cite{2010JCAP...04..014A}, the constraint is weaker
here because the subhalos inside each halo are not considered.}
(note here the exclusion limits are very conservative).
If the DS enhancement is non-negligible, the constraints will be
even more stringent. For the DS scenario with strong enhancement
($f_{\rm LW}=0,\,\tau_{\rm ds}=\infty$) and moderate enhancement
($f_{\rm LW}=10^{-4},\,\tau_{\rm ds}=\infty$), the model explaining
the $e^{\pm}$ excesses can be well excluded.

The results for $\tau^+\tau^-$ final state are shown in Fig.
\ref{fig:eg_tau}. The DM model parameters chosen for illustration are
$m_\chi=3$ TeV and $\langle\sigma v\rangle=1.9\times 10^{-22}$
cm$^{-3}$ s$^{-1}$. Similar conclusions as the $\mu^+\mu^-$ case
can be drawn.

\section{Conclusion and discussion}

In this work we study the constraints on the DM parameters using the
Fermi measured isotropic $\gamma$-ray data, taking into account the
formation of DSs accompanied with the Pop III stars. The DS formation
is expected to result in a halo with enhanced density distribution of
DM, which can give larger annihilation luminosity and produce higher
$\gamma$-ray fluxes. Two kinds of DM particle models are discussed:
the canonical neutralino-like particle, and the leptonic DM which
might be responsible for the $e^{\pm}$ excesses.

The formation rate of DSs is closely related to the SFR of Pop III
stars. We employ an analytical way to calculate
the SFR of Pop III stars. The LW feedback from the Pop III stars
themselves and the galaxies is considered. However, there is large
uncertainty of the self-shielding effect of the LW photons. A
phenomenological LW efficiency parameter $f_{\rm LW}$ is employed
to parameterize different efficiency of the LW feedback. The results
show that different $f_{\rm LW}$ can give very different SFR of
Pop III stars, and hence the enhancement effect of DM annihilation
luminosity of the DS host halos is very different. For the case
with little LW feedback ($f_{\rm LW}\lesssim 10^{-4}$) the enhancement
due to DS formation can be remarkable.

Another large uncertainty is the evolution effect of the DM halos
once DSs were formed. The halos will be subject to mergers during the
evolution of the Universe. But the change of the density profile is
not clear after the major merger of two halos with different initial
profiles. Some of the numerical simulations suggest that the resulting
density profile will be some intermediate profile after the major merger
\cite{2004MNRAS.349.1117B}, however, there are also simulations showing
the final density profile will be essentially close to the cuspier one
\cite{2006ApJ...641..647K}. Furthermore the accretion or minor merger
may also affect the evolution of the halo profile. We adopt two extreme
approaches to cover the evolution effects. One is that there is no
evolution of the DS host halos after their formation. The other is that
the DS host halos experience a fast evolution with minimum age, which
means the cuspy density profile after DS formation will disappear soon
due to one major merger. These two approaches will give very different
enhancement effects of the DS host halos. We hope further numerical
simulations will help to clarify this issue.

There are other uncertainties such as mass function, 
parameters of first stars, density profile of DM halos after
adiabatic contraction, fragmentation during the formation of the 
first stars. However, compared with the above two major uncertainties,
these uncertainties are expected to be much smaller. For example, we tested 
that if the mass function is adopted as the PS form with 
$(A,\,a,\,p)=(0.5,\,1,\,0)$ \cite{1974ApJ...187..425P}, the change of the
enhancement factor is within a factor of 2.

Given these large uncertainties, especially those from the SFR and halo 
evolution effect, our conclusion is also model dependent. We find that in 
the most optimistic case for $\gamma$-ray production through DM 
annihilation, i.e., weak LW feedback and no evolution, the DS host
halos would enhance the $\gamma$-ray signals by $1-2$ orders of
magnitude. In this case, the constraints on the cross section of the
neutralino-like DM can reach the thermal production range ($\sim 10^{-26}$
cm$^3$ s$^{-1}$). The leptonic DM models proposed to explain the
$e^{\pm}$ excesses can be well excluded, and the excess features
would be most likely of an astrophysical origin (e.g.,
\cite{2010IJMPD..19.2011F}). Conversely, if the $e^{\pm}$ excesses are
of a DM annihilation origin, then the early Universe environment should
be such that no DS can form.

In other cases with stronger LW feedback and/or minimum age of the DS
host halos, the DS enhancement is much smaller and the constraints
are much weaker. However, for the power-law model, due to the large
contribution of many low mass halos the constraints on the models to
explain $e^{\pm}$ excesses still apply.

We point out that our constraints are conservative to some extent in
several aspects. First, the constraints are derived according to the
$2\sigma$ upper bounds of the error bars of the data. Since the
observational spectrum is essentially a single power-law shape which
is very different from the DM expected spectrum, one would expect the
contribution of DM component to the EGRB would be even lower than the
observational data \cite{2010JCAP...04..014A}. Using that reduced
background would give more stringent constraints on the DS formation
scenarios and DM annihilation model. Furthermore, there are
sub-halos and sub-sub-halos in each halo, which may give even stronger
annihilation signals \cite{2010MNRAS.405..593Z}. Considering this effect 
would pose even stringent constraints. Finally, only the adiabatic 
contraction processes accompanied with the first star formation are included.
There should be many other adiabatic contraction processes, e.g. during
the formation of the second and third generation stars, which should
also give enhanced DM distribution.

\acknowledgments

We thank Kentaro Nagamine for helpful discussion on dark matter halo
evolution in the early Universe and the comments on the draft. QY
thanks Ann Zabludoff for discussion on the impact of dark star on
the first star formation rate. This work is supported by NSF under
grant AST-0908362, NASA under grants NNX10AP53G and NNX10AD48G, and
Natural Sciences Foundation of China under grant 11073024, and the
973 project under grant 2007CB815401.

\bibliographystyle{JHEP}
\bibliography{/home/yuanq/work/cygnus/tex/refs}

\end{document}